\documentclass[preprint,aps]{revtex4}
\usepackage{mathrsfs}

\usepackage{graphicx}

\begin{document}

\title{Disappearance of Nodal Gap across the Insulator-Superconductor Transition in a Copper-Oxide Superconductor}

\author{Yingying Peng$^{1}$, Jianqiao Meng$^{1}$, Daixiang Mou$^{1}$, Junfeng He$^{1}$, Lin Zhao$^{1}$, Yue Wu$^{1}$,
Guodong Liu$^{1}$,  Xiaoli Dong$^{1}$, Shaolong He$^{1}$, Jun Zhang$^{1}$, Xiaoyang Wang$^{2}$, Qinjun Peng$^{2}$,
Zhimin Wang$^{2}$, Shenjin Zhang$^{2}$, Feng Yang$^{2}$, Chuangtian Chen$^{2}$, Zuyan Xu$^{2}$, T. K. Lee$^{3}$ and X. J. Zhou$^{1,*}$
}
\affiliation{
\\$^{1}$National Laboratory for Superconductivity, Beijing National Laboratory for Condensed
Matter Physics, Institute of Physics, Chinese Academy of Sciences, Beijing 100190, China
\\$^{2}$Technical Institute of Physics and Chemistry, Chinese Academy of Sciences, Beijing 100190, China
\\$^{3}$Institute of Physics, Academia Sinica, Nankang, Taipei 11529, Taiwan
}
\date{March 05, 2013}
%
%%%\begin{abstract}

%%%\end{abstract}

\maketitle

\newpage

{\bf The parent compound of the copper-oxide high temperature superconductors is a Mott insulator. Superconductivity is realized by doping an appropriate amount of charge carriers. How a Mott insulator transforms into a superconductor is crucial in understanding the unusual physical properties of high temperature superconductors and the superconductivity mechanism. Here we report high resolution angle-resolved photoemission measurement on heavily underdoped Bi$_2$Sr$_{2-x}$La$_{x}$CuO$_{6+\delta}$ system. The electronic structure of the lightly-doped samples exhibit a number of characteristics: existence of an energy gap along the nodal direction, {\it d}-wave-like anisotropic energy gap along the underlying Fermi surface, and coexistence of a coherence peak and a broad hump in the photoemission spectra. Our results reveal a clear insulator-superconductor transition at a critical doping level of $\sim$0.10 where the nodal energy gap approaches zero, the three-dimensional antiferromagnetic order disappears, and superconductivity starts to emerge. These observations clearly signal a close connection between the nodal gap, antiferromagnetism and superconductivity.
}

\newpage

\noindent{\bf\large Introduction}

\noindent The parent compound of the copper-oxide superconductors is an antiferromagnetic Mott insulator;
doping of charge carriers into the parent compound leads to an insulator-metal transition and the
emergence of high temperature superconductivity\cite{PLee}.  In the superconducting state,
predominantly {\it d}-wave superconducting gap is well-established with a zero gap along the nodal
direction\cite{CCTsuei}. In the normal state of the underdoped superconducting region, anisotropic
pseudogap appears with a zero gap along the nodal direction\cite{TimuskReview}. A particular
yet important region exists between the parent compound and the superconducting region, i.e., a
lightly-doped region where, upon a slight doping, the three dimensional antiferromagnetic order is
rapidly suppressed and an insulator-metal transition occurs. What is the main reason for the
antiferromagnetic phase remaining insulating after being doped?  What are the electronic
characteristics of this peculiar lightly-doped region? How is the electronic structure of this
particular region connected to the pseudogap and superconducting gap in the underdoped
superconducting region? Investigation of this lightly-doped region is critical in understanding
how a Mott insulator can transform into a {\it d}-wave superconductor, the origin of the pseudogap
and its relationship with the superconducting gap\cite{IBozovic,CCOCSTM}.

In this paper we report angle-resolved photoemission (ARPES) measurements on the electronic structure
evolution with doping in Bi$_2$(Sr$_{2-x}$La$_{x}$)CuO$_{6+\delta}$ (La-Bi2201) system focusing on the
lightly-doped region and the insulator-superconductor transition. Our super-high resolution ARPES
experiments reveal an insulator-superconductor transition at a critical doping level of {\it p}$\sim$0.10.
In the lightly-doped region ({\it p}=0$-$0.10), the photoemission spectra are characterized by the coexistence
of a coherence peak and a broad hump consistent with a polaronic behavior. Moreover, a fully-opened {\it d}-wave-like anisotropic energy gap is observed which is offset by an energy gap along the (0,0)-({$\pi$,$\pi$}) nodal direction.
The nodal gap decreases with increasing doping and approaches zero at a critical doping  level of {\it p}$\sim$0.10,
where the three-dimensional (3D) antiferromagnetic order vanishes and superconductivity starts to emerge\cite{NMRGQZheng}. In addition, a smooth transition from the energy gap in the lightly-doped region to the pseudogap in the underdoped superconducting region is observed.  These observations indicate a close relationship between the nodal energy gap, antiferromagnetic insulating phase  and superconductivity. They will shed important
light on the origin of the pseudogap in high temperature superconductors, and point to the importance of
combining  the electron correlation, antiferromagnetism and strong electron-phonon coupling in understanding
the electronic structure of the lightly-doped and underdoped regions.\\

\noindent{\bf\large Results}

\noindent{\bf Transport Properties of La-Bi2201.} Figure 1 shows the temperature dependence of the in-plane resistivity $\rho_{ab}$ (Fig. 1a) and the
magnetization (Fig. 1b) of the La-Bi2201 single crystals. The La-Bi2201 samples with {\it p}=0$\sim$0.08
doping levels are non-superconducting. The {\it p}=0.10 sample is superconducting with a {\it T}$_{\rm c}$ at $\sim$3 K,
which is on the border of insulator-to-superconductor transition. The {\it p}=0.105 sample has a {\it T}$_{\rm c}$ of 12 K.

It is interesting to note that, the insulator-superconductor transition takes place near the
universal two-dimensional resistance, {\it h}/4$e^2$, which corresponds to three-dimensional resistivity
around 0.8 m$\Omega$$\cdot$cm in Bi2201. The insulator-superconductor transition can also be driven by Zn
doping\cite{Fukuzumi49} or by changing the thickness of cuprate thin films\cite{IBozovic}, and it is known that it takes
place at nearly the same resistance values.

The in-plane resistivity-temperature curves of the La-Bi2201 single crystals with low doping levels
of 0.03, 0.04 and 0.08 are re-plotted in Fig. 1c as ln$\rho$$\sim$$T^{-1/3}$. Over an extended range of low temperature,
these curves show a linear dependence. This is consistent with the two-dimensional variable range hopping
model where the resistivity-temperature dependence is expected to be: $\rho$=$\rho_0$$e^{{({T_0}/T)}^{1/3}}$\cite{Mott50,Fiory51}.

\noindent{\bf Underlying Fermi surface.} Figure 2 shows the doping evolution of the underlying Fermi surface across the non-superconductor-superconductor
transition and indications of the nodal energy gap observed for the lightly-doped La-Bi2201 samples.
As seen from Fig. 2(a-d), in the lightly-doped samples, the spectral weight distribution dominates
near the ($\pi/2$,$\pi/2$) nodal region and then spreads towards the (0,$\pi$) anti-nodal region
with increasing doping. As commonly used before, we denote the high intensity contour in the spectral
weight distribution as the ``underlying Fermi surface", although in principle a strict
Fermi surface refers to the locus of momentum without an energy gap\cite{UnderlyFS}. The photoemission
spectra (energy distribution curves, EDCs) along the underlying Fermi surface are displayed in Fig. 2(e, g, i, k) for
different doping levels, with their corresponding symmetrized EDCs shown in Fig. 2(f, h, j and l),
respectively. The EDC symmetrization is an empirical and intuitive way to examine the existence of an
energy gap by removing the Fermi distribution function from the original ARPES data\cite{NormanEDC}.
As seen from Fig. 2e and 2f, with very low doping of {\it p}=0.03, only a broad hump feature dominates
the measured spectra. It disperses toward low binding energy, reaches its minimum around 200 meV along
the nodal direction and then turns back. The spectral weight near the Fermi level {\it E}$_{\rm F}$ is strongly
suppressed for this lightly-doped sample and there is apparently a gap opening.  Further increase of
the doping to  {\it p}=0.055 results in an emergence of a coherence peak near the Fermi level which
coexists with the broad hump structure (Fig. 2g). A close inspection reveals an energy gap for the
coherence peak along the entire underlying Fermi surface, including the nodal region (Fig. 2h).
When the doping level reaches {\it p}=0.08, the coherence peak becomes prominent. In the mean time,
the broad hump gets weaker but remains visible (Fig. 2i). The energy gap opening is clear along the
entire underlying Fermi surface(Fig. 2j).  Eventually, when the doping level becomes {\it p}=0.10,
the coherence peak gets sharp and the broad hump remains discernable (Fig. 2k). But in this case,
no indication of an energy gap opening is observed near the nodal region (Fig. 2l).  However,
there are clear indications of gap opening in the EDCs far away from the nodal region. This is
consistent with the observation of the Fermi arc in underdoped superconducting samples\cite{NormanArc,KanigelNodalLiquid,WSLee}.

\noindent{\bf Nodal band structure.} Figure 3 focuses on the nodal gap and its detailed doping evolution. Fig. 3(a-g) shows doping
evolution of the band structure along the nodal direction. The Fermi distribution function is
divided out from these data to facilitate the revelation of the energy gap near the Fermi level.
Original EDCs for the La-Bi2201 samples with different doping levels  are shown in Fig. 3h on
the underlying Fermi surface along the nodal direction, with their corresponding symmetrized EDCs
shown in Fig. 3i. From the raw data (Fig. 3(a-g)), it is clear that, with increasing doping, the
spectral weight transfers from high binding energy to low binding energy. For the {\it p}=0.03 sample,
the spectral weight near {\it E}$_{\rm F}$ is strongly depleted (Fig. 3a) and the nodal EDC is characterized by a
prominent broad hump near 250 meV.  A slight increase of the doping level to 0.04 brings a significant
amount of spectral weight towards low binding energy and a weak coherence peak emerges near the Fermi
level (Fig. 3b and 3h). Further doping brings more spectral weight to the Fermi level so that the
coherence peak becomes sharper while the broad hump structure becomes weaker.  The relative spectral
weight increase of the coherence peak with increasing doping is shown in the inset of Fig. 3h (red circles).
Similar trend was observed before in other cuprates\cite{TYoshida,KShenPRL}. We note that, compared with the
previous results\cite{HashimotoPRB,HDingPRB},  this is the first time one can observe clear coherence peaks
below the doping level of {\it p}=0.10 in Bi2201 and a coexistence of peak-dip-hump structure for such low
doping samples. This is mainly due to sample quality improvement and the utilization of super-high resolution
of our laser ARPES measurements. The observation of a clear coherence peak near the Fermi level makes it
possible to precisely determine the energy gap. As shown in the symmetrized nodal EDCs in Fig. 3i,
whereas the symmetrized EDC of the {\it p}=0.03 sample becomes featureless near {\it E}$_{\rm F}$, there is
a clear indication of nodal gap opening in the samples with doping of {\it p}=0.04 to {\it p}=0.08.
Such a nodal gap is closed for the {\it p}=0.10 and {\it p}=0.105 samples. As shown in the inset
of Fig. 3h, the nodal gap drops monotonically with increasing doping till the doping level of 0.10
which just becomes superconducting with a {\it T}$_{\rm c}$ around 3 K.

In order to understand the origin of the nodal gap in the lightly doped samples, it is important to
investigate its temperature dependence. Fig. 4 shows the temperature evolution of the band structure
and the nodal gap in a typical La-Bi2201 sample with a doping level of 0.055. The measured data  are
divided by the Fermi distribution function (Fig. 4a) in order to visually inspect the energy gap as
well as the possible features above the Fermi level. There is clearly a spectral weight suppression
near the Fermi level for all the temperatures we have covered. This indicates that the nodal gap
persists in the entire temperature range up to 300 K, close to the temperature of the possible pseudogap
as determined by the NMR meausrements\cite{NMRGQZheng}. From the original EDCs on the underlying Fermi
surface along the nodal direction (Fig. 4b),  one can see that, with increasing temperature, the
coherence peak gets weaker and becomes indiscernible above 150 K, while the high energy hump shows
a slight change with temperature.  The corresponding EDCs after dividing out the Fermi distribution
functions (Fig. 4c) exhibit a dip near the Fermi level, which together with the spectral distribution
in the original data (Fig. 4a), indicate that the gap below and above the Fermi level is nearly
symmetric. This justifies the procedure in symmetrizing EDCs to extract the energy gap  in the
paper. Fig. 4d shows symmetrized EDCs on the underlying Fermi surface along the nodal direction;
one can also see the persistence of the nodal gap till high temperatures. We do not show EDCs
above 150 K in Fig. 4(b-d) because the disappearance of the coherence peak at high temperatures
makes it difficult to quantitatively determine the gap size.

\noindent{\bf Momentum dependence of gap.} To investigate the momentum dependence of the energy gap in the lightly-doped La-Bi2201,
we have measured various momentum cuts along the underlying Fermi surface, picked up the EDCs
along the underlying Fermi surface like in Fig. 2(e, g, i ,k), and extracted the gap size from
the peak position of the symmetrized EDCs like in Fig. 2(f, h, j, l). The obtained gap size as a
function of momentum is shown in Fig. 5(a-e) for the La-Bi2201 samples with doping
levels at 0.04, 0.055, 0.07, 0.08 and 0.10, respectively.  It is clear that the gap is
anisotropic and increases when the momentum moves from the nodal to the antinodal regions.
For comparison, we also plotted an offset {\it d}-wave-like gap form $\Delta$=$\Delta_0$cos(2{\it $\Phi$})+$\Delta_{\rm N}$ with $\Delta_{\rm N}$
representing the nodal gap (solid dashed lines in Fig. 5(a-d)) and one can see that the
measured data basically follow such a simple form in the samples with doping {\it p}=0.04$\sim$0.08.
This is consistent with the previous report that {\it d}-wave like gap is observed in the parent
compound Ca$_2$CuO$_2$Cl$_2$\cite{FRonning}, insulating Bi$_2$Sr$_2$CaCu$_2$O$_{8+\delta}$ (Bi2212)\cite{Chatterjee},
heavily underdoped Bi2212\cite{IVishik} and underdoped (La$_{2-x}$Sr$_x$)CuO$_4$\cite{LSCONGap}.
The presence of the nodal gap indicates that the underlying Fermi surface is fully gapped for these samples.
At the critical doping {\it p}$\sim$0.10 where superconductivity emerges, it shows zero gap near the nodal
region which is consistent with the ``Fermi arc" picture in the pseudogap state of other copper-oxide
superconductors\cite{NormanArc,KanigelNodalLiquid}.  Fig. 5(f-k) shows the position of the
high energy hump as a function of momentum for the La-Bi2201 samples at different doping levels.
The broad hump structure also exhibits anisotropic momentum dependence with its position following
a form $P_{\rm {h}}$=$P_{\rm {h0}}$cos(2$\Phi$)+$P_{\rm {hN}}$ with $P_{\rm {hN}}$ representing the hump position along the
nodal direction (Fig. 5(f-k)). The hump structure exhibits a minimum in binding energy along the nodal
direction and its position decreases with the increase of doping levels. The similar momentum dependence
of both the energy gap and the hump, and their similar doping dependence, suggest that they are
intimately related, as will be discussed more below. Fig. 5l illustrates the evolution
of the energy gap and its relation with the three-dimensional antiferromagnetism (AF)
and superconductivity\cite{NMRGQZheng}. At a critical doping level of {\it p}$\sim$0.10,
the nodal gap approaches zero, the three dimensional antiferromagnetism vanishes, and superconductivity (SC) starts to
emerge.\\

\noindent{\bf\large Discussion}

\noindent Our observation of the nodal gap, and hence a fully-gapped underlying Fermi surface in the
lightly-doped La-Bi2201 samples ({\it p}=0$-$0.10) indicates they have an insulating ground
state. This is in a good correspondence to the transport measurements where the insulating
behavior is observed until {\it p}=0.08 in the resistivity-temperature dependence (Fig. 1a).
Previous optical measurements also indicated a gap opening with the transition from metal to insulator near
a doping of {\it p}$\sim$0.10 in La-Bi2201\cite{SLupi}. The temperature dependence of the resistivity fits well
with the model of two-dimensional variable range hopping of localized charges (Fig. 1c).
For lightly-doped samples, there is a semiconducting or insulating behavior observed in the resistivity-temperature
dependence\cite{HTakagi,YAndo} but it is unclear whether it is a character of an intrinsic band insulator or a charge
localization effect. Our results favor the latter. Our work clearly indicates that there is a true gap opening or
complete loss of spectral weight at the Fermi level along the entire underlying Fermi surface so that the lightly-doped samples are insulators.

As shown in Ref.\cite{UnderlyFS}, the high intensity contour in the spectral weight distribution corresponds to the underlying Fermi
surface only when the gap is small. Since we focus mainly on the nodal region where the gap is relatively small, the maximal intensity
contour is close to the underlying Fermi surface and the gap we have observed represents the one that opens up on top of the Fermi
surface because of correlation. We note that, while the nodal gap has been reported before in a number of underdoped
cuprates\cite{KShenNGap,IVishik,LSCONGap}, our observations are different in an important aspect. It was reported that a
nodal gap is observed in the underdoped but superconducting Bi2212 above {\it T}$_{\rm c}$\cite{IVishik} although there
were different reports that no nodal gap is present even in a highly underdoped insulating Bi2212 sample\cite{Chatterjee,HBYang}.
For the underdoped Bi2212 sample with  a {\it T}$_{\rm c}$ near 34 K which was reported to exhibit a nodal energy gap, its
resistivity-temperature dependence shows a good metallic behavior over an entire temperature range\cite{XFSun}.
In (La$_{2-x}$Sr$_x$)CuO$_4$ system, a nodal gap was reported at a doping level of 0.08 with a {\it T}$_{\rm c}$ at 20 K\cite{LSCONGap},
although transport measurements also indicate that it is already a good metal over a whole temperature range\cite{HTakagi,YAndo}.
These results suggest that the nodal gap, the pseudogap and the superconducting gap can occur at the same doping
level\cite{IVishik,LSCONGap}. In our case of La-Bi2201, it is clear that the nodal gap is competing with the superconducting gap: superconductivity
starts to emerge only after the nodal gap disappears. We note that the super-high resolution ARPES measurements enable
us to observe clear coherence peaks even in the heavily underdoped samples, thus making it possible to get a precise
determination of energy gap.  Also, for many samples investigated in the present study, they have the same composition;
the doping is slightly tuned by varying the oxygen content during the annealing process.

The existence of the nodal gap in the lightly-doped La-Bi2201 samples ({\it p}$<$0.10), together with other characteristics
like the peculiar doping, momentum and temperature dependence, puts constraints on its underlying origin of formation. The
coincidence of the nodal gap disappearance and the vanishing of the three-dimensional antiferromagnetism near {\it p}$\sim$0.10
indicates that long-range antiferromagnetism is closely related to the electronic structure evolution across the insulator-superconductor
transition. It is natural to ask whether the static antiferromagnetic order alone can give rise to the energy gap formation in the
lightly-doped region. As it is well-known, the presence of static antiferromagnetic order can double the unit cell in real space and
reduce the Brillouin zone in the reciprocal space by half. This would cause band folding that will make the initial ``large Fermi surface"
and folded Fermi surface cross somewhere in between the nodal and antinodal regions. This is most clearly demonstrated in the electron-doped
cuprates\cite{HMatsui}. This picture alone cannot explain the gap formation near the nodal region, it cannot explain the anisotropic
{\it d}-wave-like gap form either. Very recently it was pointed out by theoretical calculations that the nodal {\it d}-wave spectrum
is robust against strongly fluctuating competing order like antiferromagnetic correlation.  But when the antiferromagnetic correlation
is strong and the phase has spatial inhomogeneity, a nodal gap can be developed which grows with the antiferromagnetic magnitude\cite{Atkinson}.
This picture may be able to account for the nodal gap opening in the lightly-doped region. However, the result is obtained in a weak
coupling approach and it shows a strong temperature dependence that is different from what we have observed.

Another possibility to consider is whether the energy gap can be induced by disorder because it can give rise to localization which
leads to insulating behaviors. In doped semiconductors which show variable range hopping, there is a long range Coulomb effect that
can depress the density of states near the Fermi energy forming the
so-called ``Coulomb gap"\cite{ESEffect}. Since our La-Bi2201 samples ({\it p}$<$0.10) are insulating at low temperature with its
transport property consistent with the variable range hopping picture (Fig. 1c), this effect may be present. The
disorder scenario can in principle account for our observation of the nodal gap; it was proposed to explain the normal-state
antinodal gap before\cite{ZHPan}. In this picture, it is also possible to explain the nodal gap increase with decreasing doping
because the screening of the Coulomb interaction gets weaker. However, it is unclear whether such an interaction effect can produce
a {\it d}-wave-like anisotropic energy gap and why the disappearance of the nodal gap can coincide with the vanishing of the
long-range antiferromagnetism. The presence of the high energy hump does not have a natural explanation from the disorder effect
either. One may invoke strong electron correlation that can generate a hump structure as the incoherent background which lies
below the coherent states near the Fermi level\cite{XYZhang}. It has been recently shown theoretically that disorder can drive
an insulator-superconductor transition at a critical disorder strength and produce a single particle gap\cite{DisorderNP}. These
aspects bear similarities to our observations of the nodal energy gap and an insulator-superconductor transition near {\it p}$\sim$0.10.
However, it predicts that the nodal gap exists for both the insulating region and the superconducting region, which is inconsistent
with our observation that the nodal gap exists only in the insulating samples.

It is known that in the heavily underdoped regime, charge ordering like the diagonal stripe may be formed, as observed in
La$_{2-x}$Sr$_x$CuO$_4$ (LSCO) system\cite{LSCOstripe1,LSCOstripe2} as well as Bi$_{2-x}$Sr$_{2+x}$CuO$_{6+y}$ (Bi2201)
system\cite{Bi2201stripe}.  It was suggested that the diagonal stripe can be understood in terms of the conventional
spin-density wave (SDW) picture, originating from nesting of nodal states\cite{Valla2012}. Such SDW is expected to open
a gap and localize the nodal states that prevent the superconductivity from occurring. When the diagonal SDW vanishes,
the gap closes, the nodal states are recovered and superconductivity emerges. These appear to agree with our results that
the nodal energy gap approaches zero when the superconductivity emerges. However, from the diagonal stripe scenario, it
is not clear why the disappearance of the nodal gap coincides with the vanishing of the long range antiferromagnetism.
It is also unclear why the gap takes an anisotropic {\it d}-wave-like gap form since the nesting order would open a gap
only on the nested segments of the Fermi surface\cite{Vallascience}.

The systematic doping evolution of the photoemission spectra, i.e., the emergence of the coherence peak with a slight
doping, the coexistence of the coherence peak with a broad hump, and their coordinated evolution with doping (Fig. 3h and Fig. 5),
is reminiscent of observations in other copper-oxide compounds\cite{TYoshida,KShenPRL}, some one-dimensional
materials\cite{PerfettiPolaron} and bilayer manganite La$_{1.2}$Sr$_{1.8}$Mn$_2$O$_7$\cite{Mannella}. These behaviors
can be reasonably explained by the formation of small polarons\cite{PerfettiPolaron,KShenPRL,ORosch,XJZhouReview} with
electrons dressed by spin excitations or lattice vibrations. This polaron picture can explain the lineshape of the photoemission
spectra\cite{PerfettiPolaron,KShenPRL,ORosch,XJZhouReview}. However, it does not provide a satisfactory explanation for the
energy gap because, in a simple polaron picture, the primary coherence peak is expected to stay at the Fermi level\cite{PerfettiPolaron}.
Theoretical calculations\cite{Sangiovanni,Mishchenko} have shown that antiferromagnetism can enhance the effective electron-phonon
interaction that makes it possible for the formation of the self-trapped polarons.  Once the self-trapped polaron is formed, the
spectral weight of the quasiparticle at the Fermi level is reduced to zero\cite{Mishchenko}. In addition, there are several
peaks appeared at higher binding energies as the motion of the charge is strongly coupled with the creation of spin wave
excitations and phonons.  Therefore, in the extremely low doping region (like {\it p}=0.03 and below) with a strong
antiferromagnetic order, the effective electron-phonon interaction is very strong, those peaks appeared at rather
high binding energy without spectral weight at the Fermi level, leading to the appearance of a large gap. With increasing
doping (like {\it p}=0.04$\sim$0.08), the peaks would move towards the Fermi level, accompanied by a decrease of the gap size.
When the effective electron-phonon interaction reduces below a critical value, the gap approaches zero and the electron-phonon
interaction is simply to renormalize the quasiparticle mass with a finite spectral weight at the Fermi level. Our observation
of the simultaneous disappearance of the antiferromagnetism and the nodal gap (or lack of spectral weight at the Fermi level)
seems to agree quite well with this picture.  One important prediction of the theories\cite{ Mishchenko,Bonca} is that the
momentum dependence of the peaks at higher binding energies follows the same energy dispersion of the quasiparticles without
any electron-phonon interaction. Thus the {\it d}-wave like momentum dependence of the ARPES spectra shown in Fig. 5 can be
naturally understood. We also note that Fig. 5(a-e) shows a rigid energy shift along almost the entire Fermi surface. Any
mechanism for the formation of the nodal gap should not modify the Fermi surface and the polaron picture is a natural
candidate for this effect. Although the polaron scenario seems quite plausible for our observations, we still need a more
quantitative explanation of the spectral lineshape for the insulating regime. To this end, a more comprehensive calculation
may be necessary to include both the electron-phonon interaction and long-range Coulomb effect together with the antiferromagnetism.

One important question to ask is the relationship between the energy gap observed in the insulating
samples ({\it x}=0$\sim$0.08) and the pseudogap observed in the underdoped superconducting region ({\it x}=0.10$\sim$0.15).
In Fig. 5l, for several given momenta, we plotted the energy gap observed in the insulating region ({\it x}=0$\sim$0.08)
and the pseudogap observed in the underdoped superconducting region ({\it x}=0.10$\sim$0.105). The region with the
insulating gap smoothly evolves into the pseudogap region without an abrupt change. In conjunction with
the {\it d}-wave-like gap form, the decrease of the nodal gap as doping increases from 0.04 to 0.10 also
reduces the gap value at the antinodal region to about 40 meV which is close to the maximum pseudogap value
observed for the superconducting phase\cite{Chatterjee}. This is the first experimental evidence that has
been presented about the continuous increase  of the pseudogap at the antinodal region as the system crosses
from the superconducting phase to the insulating antiferromagnetic phase. Our result of the {\it d}-wave-like gap form
is qualitatively consistent with the idea that the {\it d}-wave-like spin singlet pairing already exists in the insulating
parent compound and lightly-doped samples\cite{RVBTheory}. It is theoretically shown that, even if there is a long
range antiferromagnetic order, energy dispersion still has the {\it d}-wave-like form along the Fermi surface\cite{Lee}.
However, due to the presence of the antiferromagnetic phase, the electron-phonon interaction is greatly enhanced to
create polarons and needs to be considered to account for the spectral lineshape and the nodal gaps we have observed.
Since the antiferromagnetic fluctuation and electron-phonon coupling are present well into the higher doping region
in spite of their reduced strength\cite{XJZhouReview}, one may expect they also play important role in dictating
the electronic structure and properties in higher doping regions.

%In summary, we have observed a clear insulator-superconductor transition in the La-Bi2201 system near the doping
%level of $\sim$0.10, taking advantage of our super-high resolution laser ARPES measurements.   The nodal gap gets small with the increase of the doping level and approaches zero at the critical doping {\it p}$\sim$0.10 where the three-dimensional antiferromagnetic order vanishes and superconductivity starts to emerge. These observations are qualitatively consistent with the formation of self-trapped polarons in the lightly-doped samples. They indicate that antiferromagnetic correlation and the strong electron-phonon interaction are important to understand the electronic structure and  anomalous physical properties of cuprates, at least in the underdoped region.\\

In conclusion, we have observed a clear insulator-superconductor transition in the La-Bi2201 system near the doping
level of $\sim$0.10. The results reveal an intimate relation between superconductivity and antiferromagnetism, i.e., superconductivity
emerges at the critical doping level where three-dimensional antiferromagnetic order vanishes and the nodal gap
disappears. These observations are qualitatively consistent with the formation of self-trapped polarons in the lightly-doped samples.
They indicate that antiferromagnetic correlation and the strong electron-phonon interaction are important to understand
the electronic structure and anomalous physical properties of cuprates, at least in the underdoped region.\\

\noindent{\bf\large\textbf{Methods} }

\noindent{\bf Sample preparation and estimation of doping levels.} The Bi$_2$Sr$_{2-x}$La$_{x}$CuO$_{6+\delta}$ (La-Bi2201) single crystals were grown by the traveling solvent floating
zone method\cite{JQMengCrystal}. We obtained various dopings by changing the La concentration and post-annealing under different temperatures and
atmospheres: the sample with the hole concentration {\it p}=0.03 has a composition of {\it x}=1.02 while all the other samples with
hole concentrations of 0.04, 0.055, 0.07, 0.08, 0.10 and 0.105 have the same composition with {\it x}=0.84
but annealed under different conditions.  The La-Bi2201 samples are characterized by different measurements, including magnetization measurements,
the temperature dependence of the resistivity and the Hall coefficient measurements. The doping level {\it p} in La-Bi2201 is
determined by a couple of factors including La-content, oxygen content, inter-exchange between Bi and Sr and so on. Because of these complications,
the precise determination of the doping level {\it p} is not straightforward but was estimated in the literatures from combined transport measurements
like resistivity, Hall coefficient, and the thermoelectric power. In our present work, we followed similar procedure to estimate the doping levels
by comparing our data with those in literature that already provided a good correspondence between the doping concentration, {\it T}$_{\rm c}$ and resistivity-temperature
dependence\cite{OnoPRB}.  Note that we used similar single crystal growth procedures as used by other groups\cite{NMRGQZheng,OnoPRB} so the doping levels between the samples are
presumably comparable. The La-Bi2201 samples with hole concentrations {\it p} of 0.03,
0.04, 0.055, 0.07 and 0.08 are non-superconducting. The {\it p}=0.10 sample is superconducting with a
transition temperature {\it T}$_{\rm c}$$\sim$3 K, near the border of nonsuperconductor-to-superconductor
transition. The {\it p}=0.105 sample shows a {\it T}$_{\rm c}$ of 12 K.\\

\noindent{\bf High resolution ARPES measurements.} The ARPES measurements were carried out on our vacuum ultraviolet (VUV) laser-based angle-resolved
photoemission system with advantages of high photon flux, enhanced bulk sensitivity and super-high energy and momentum resolution\cite{GDLiu}.
The photon energy is 6.994 eV with a bandwidth of 0.26 meV. We set the energy resolution of the electron energy analyzer (Scienta R4000) at 1.5 meV,
giving rise to an overall instrumental energy resolution of 1.52 meV. The angular resolution is $\sim$0.3$^{\circ}$, corresponding to a momentum
resolution of $\sim$0.004${\AA}$$^{-1}$. The samples were cleaved \emph{in situ} and measured under ultrahigh vacuum better than 4$\times$10$^{-11}$ Torr.\\

$^{*}$Correspondence and requests for materials should be addressed to X.J.Z. (XJZhou@aphy.iphy.ac.cn).\\

\noindent{\bf\large

\vspace{3mm}

\noindent {\bf\large Acknowledgement}\\
We acknowledge helpful discussions with Chandra Varma, Maurice Rice, Tao Xiang and Zhengyu Weng.  XJZ thanks the funding support from NSFC (Grant No. 11190022) and the MOST of China (Program No: 2011CB921703 and 2011CB605903).

\vspace{3mm}

\noindent {\bf\large Author Contributions}\\
 X.J.Z. and Y.Y.P. proposed and designed the research. Y.Y.P., J.Q.M., D.X.M., J.F.H., L.Z., G.D.L., X.L.D., S.L.H., J.Z., X.Y.W., Q.J.P., Z.M.W., S.J.Z., F.W., C.T.C., Z.Y.X. and X.J.Z contributed to the development and maintenance of Laser-ARPES system. Y.Y.P. prepared and characterized the samples with help from J.Q.M., Y.W. and X.L.D.. Y.Y.P. carried out the experiment and data analysis with the assistance from J.Q.M., D.X.M., J.F.H., L.Z., G.D.L., S.L.H. and X.J.Z.. X.J.Z., Y.Y.P. and T.K.L. wrote the paper.

\vspace{3mm}

\noindent{\bf Competing financial interests:} The authors declare no competing financial interests.

\newpage

\begin{figure}[tbp]
\begin{center}
\includegraphics[width=1\columnwidth,angle=0]{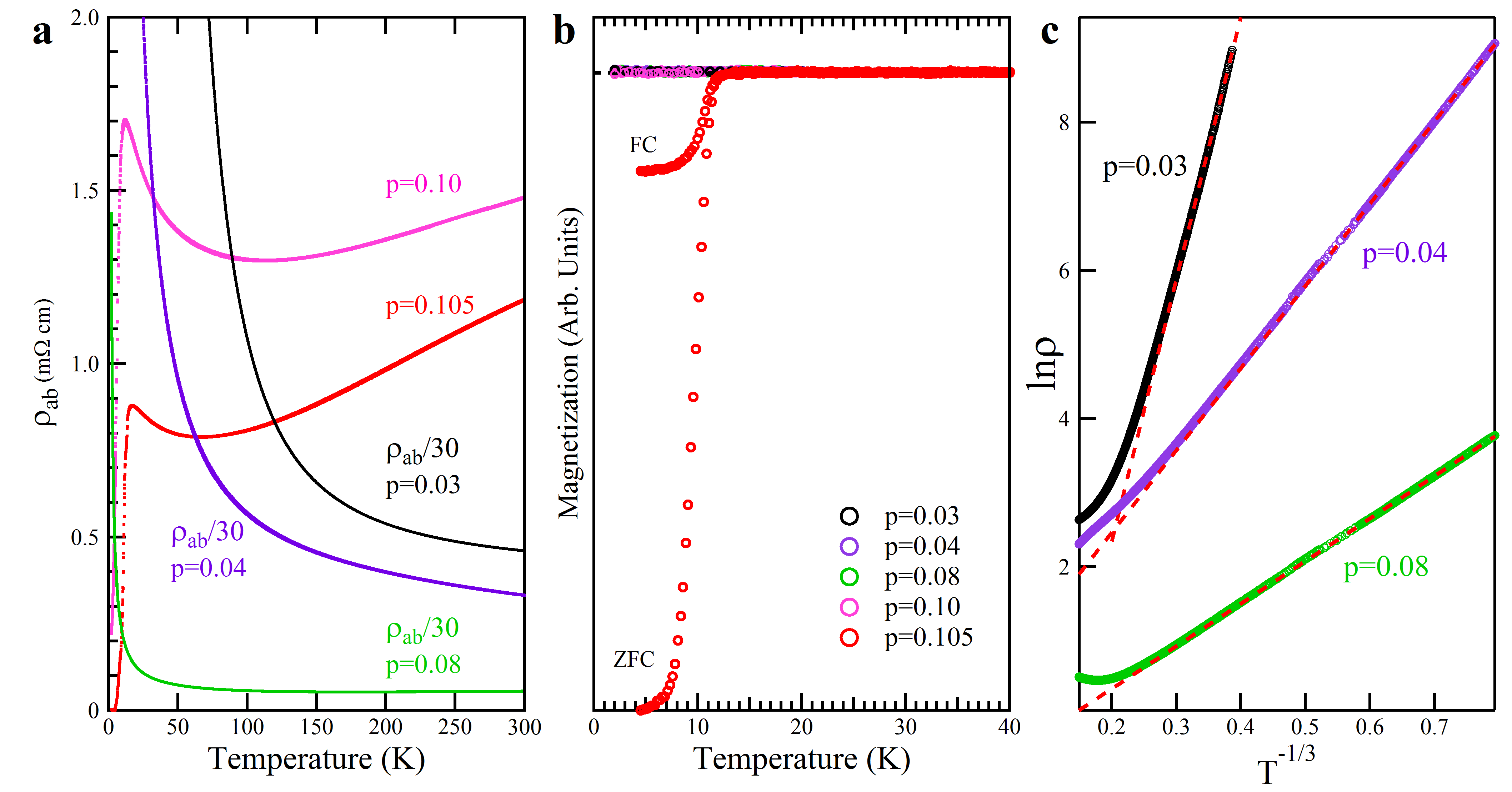}
\end{center}
\caption{{\bf Transport and magnetic measurements of the La-Bi2201 samples.} ({\bf a}) Temperature dependence of the in-plane resistivity $\rho_{ab}$ of the
La-Bi2201 single crystals. ({\bf b}) Magnetization measurements of the La-Bi2201 single crystals with a magnetic field of 1 Oe. FC refers to
the field-cooled measurement while ZFC refers to zero field cooled measurement. ({\bf c}) The in-plane resistivity-temperature curves plotted as ln$\rho$$\sim$$T^{-1/3}$
for the La-Bi2201 single crystals with low doping levels of 0.03 (black circles), 0.04 (blue circles) and 0.08 (green circles). The dashed red lines are
linear that serve as guide to the eyes. Over an extended range of low temperature, these curves show a linear dependence.}
\end{figure}

\begin{figure}[tbp]
\begin{center}
\includegraphics[width=1\columnwidth,angle=0]{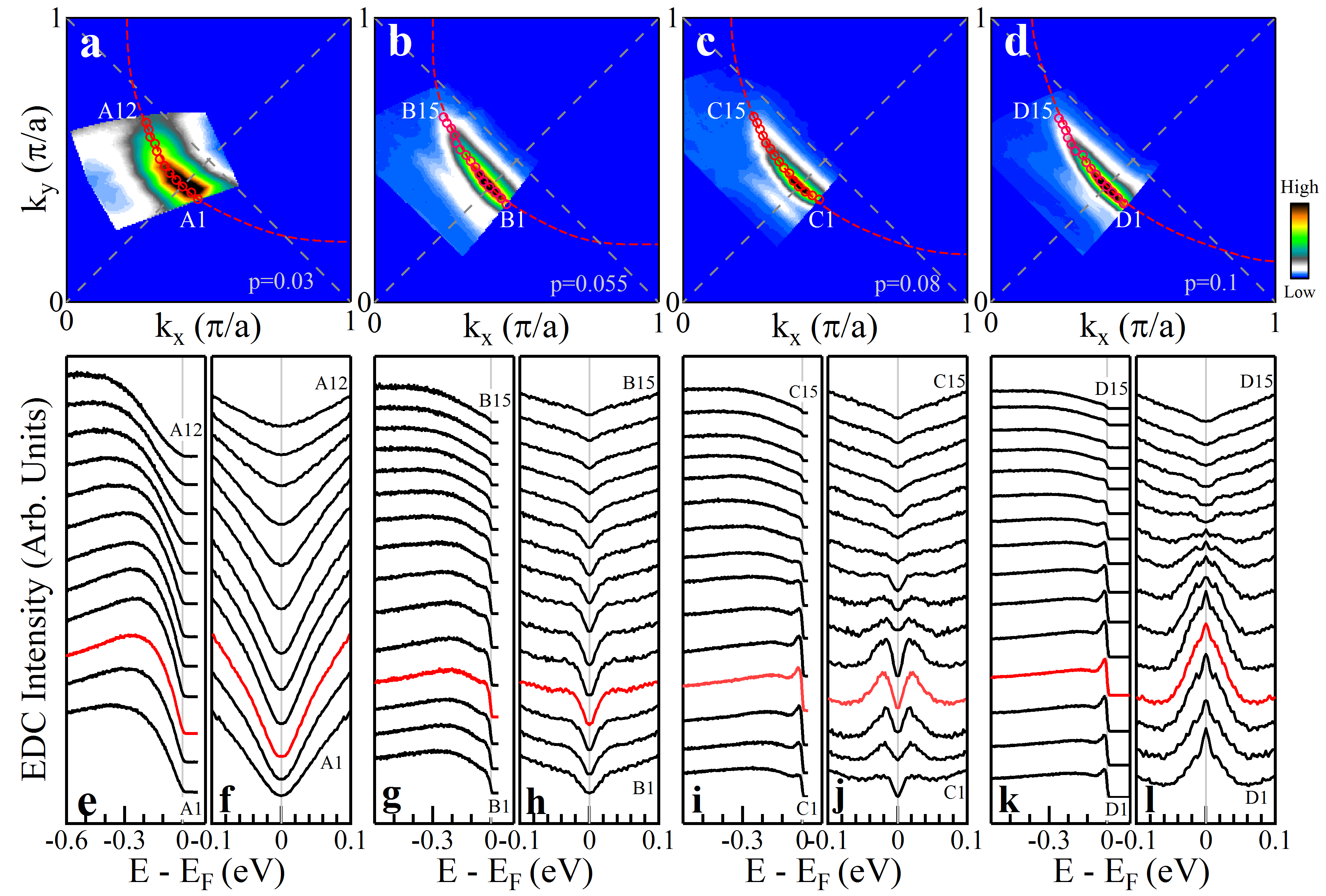}
\end{center}
\caption{{\bf Observation of the nodal gap in lightly doped La-Bi2201 samples.} ({\bf a} to {\bf d}) The momentum distribution of the spectral weight
integrated within a [-10meV,10meV] energy window around {\it E}$_{\rm F}$ for La-Bi2201 samples with different doping levels {\it p}
of 0.03 ({\bf a}), 0.055 ({\bf b}), 0.08 ({\bf c}) and 0.10 ({\bf d}). The data were taken at a temperature of 20 K. The location of
the ``underlying Fermi surface" is labeled by red empty circles and the red dashed lines are obtained by
tight-binding fitting to the circles which serve as a guide to the eyes. The photoemission spectra (energy distribution curves, EDCs)
along the ``underlying Fermi surface" are shown in ({\bf e}), ({\bf g}), ({\bf i}) and ({\bf k}) for the doping levels of 0.03 ({\bf e}),
0.055 ({\bf g}), 0.08 ({\bf i}) and 0.10 ({\bf k}), respectively. The corresponding symmetrized EDCs are shown in ({\bf f}), ({\bf h}),
({\bf j}) and ({\bf l}). Red lines indicate the EDCs and symmetrized EDCs along the (0,0)-($\pi$,$\pi$) nodal direction. We paid special
attention to avoid any charging effect due to the insulating behaviors of these lightly-doped samples. We varied the photon flux over a
large range and observed little change in the measured EDCs.}
\end{figure}

%%\begin{figure*}[floatfix]
\begin{figure}[tbp]
\begin{center}
\includegraphics[width=1\columnwidth,angle=0]{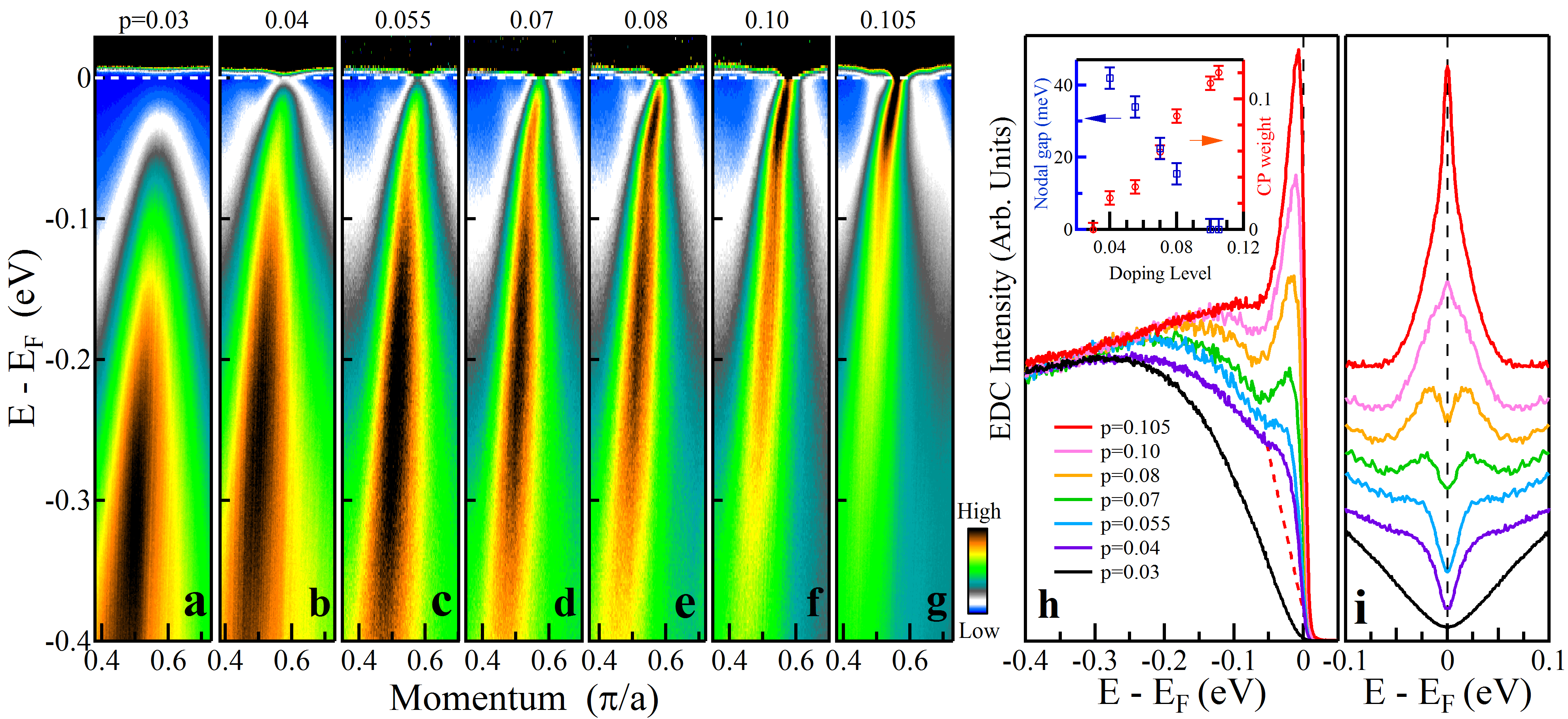}
\end{center}
\caption{{\bf Doping evolution of the nodal gap in lightly-doped La-Bi2201 samples.} ({\bf a} to {\bf g}) Band structures measured along the nodal
direction for the La-Bi2201 samples with different doping levels {\it p} of 0.03 ({\bf a}), 0.04 ({\bf b}), 0.055 ({\bf c}),
0.07 ({\bf d}), 0.08 ({\bf e}), 0.10 ({\bf f}) and 0.105 ({\bf g}). The data were taken at a temperature of $\sim$15 K and were
divided by the Fermi distribution function to highlight the gap opening near the Fermi level. ({\bf h}) shows the original EDCs on
the ``underlying Fermi surface" along the nodal direction; the corresponding symmetrized EDCs are shown in ({\bf i}). The top-left
inset in ({\bf h}) shows the doping dependence of the nodal gap (blue empty squares) and the coherence peak (CP) spectral weight
(red empty circles). The nodal gap is obtained by taking the peak position of the symmetrized EDCs in ({\bf i}). The coherence peak
spectral weight (CP weight) is defined by the ratio between the peak area and the overall spectral weight integrated over
the [-0.4eV,0.03eV] energy window.  The coherence peak area is obtained by subtracting a linear background (red dashed line) as illustrated in ({\bf h}) for the EDC of the {\it p}=0.04 sample.
}
%%\end{figure*}
\end{figure}

\begin{figure}[tbp]
\begin{center}
\includegraphics[width=1.0\columnwidth,angle=0]{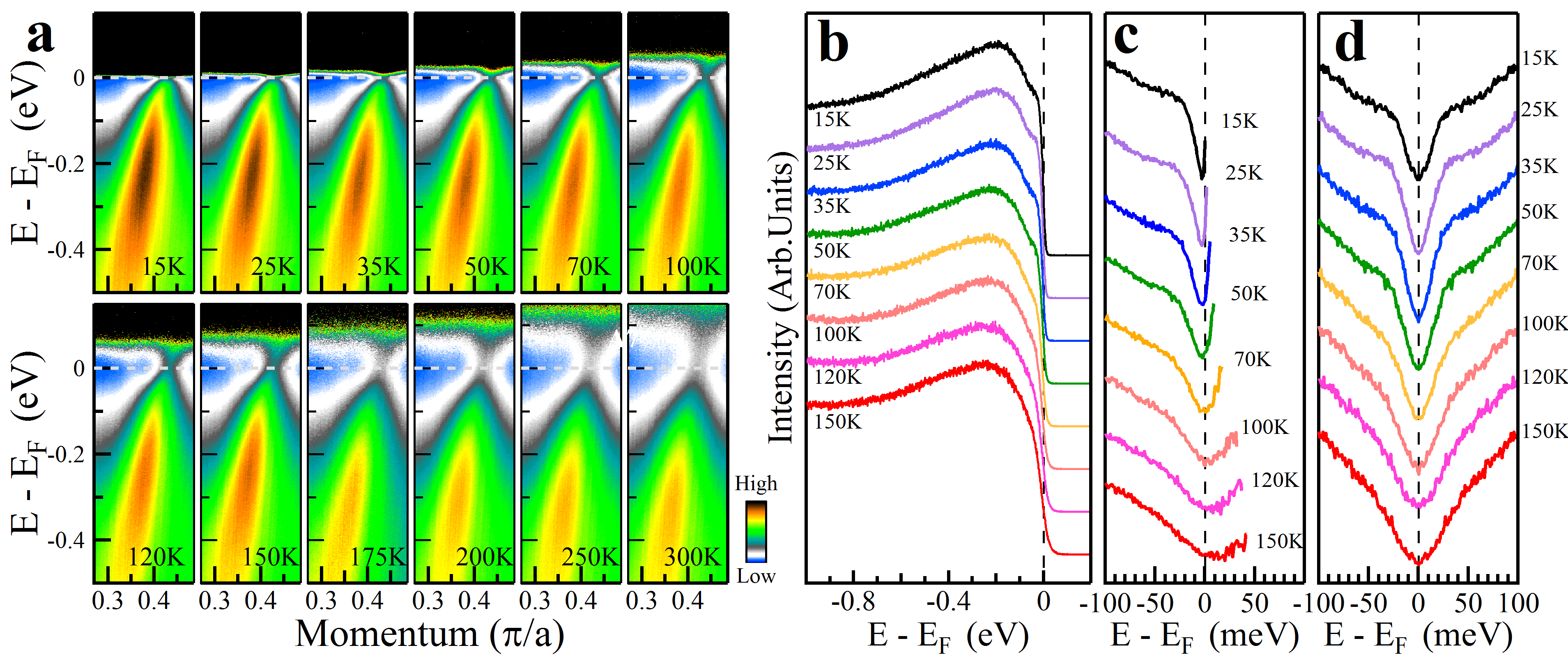}
\end{center}
\caption{{\bf Temperature dependence of the nodal gap.} ({\bf a}) Band structure of the La-Bi2201 sample with a doping level of {\it p}=0.055
measured along the nodal direction at different temperatures. The data have been divided by the corresponding Fermi distribution functions to
highlight the energy gap and the features above the Fermi level. ({\bf b}) Original EDCs at the underlying Fermi momentum at different
temperatures. ({\bf c}) Corresponding EDCs after being divided by the Fermi distribution functions. ({\bf d}) Symmetrized EDCs obtained
from ({\bf b}). Sample ageing has been checked by cycling temperature and the data are reproducible during the process.
}
\end{figure}

\begin{figure}[tbp]
\begin{center}
\includegraphics[width=1.0\columnwidth,angle=0]{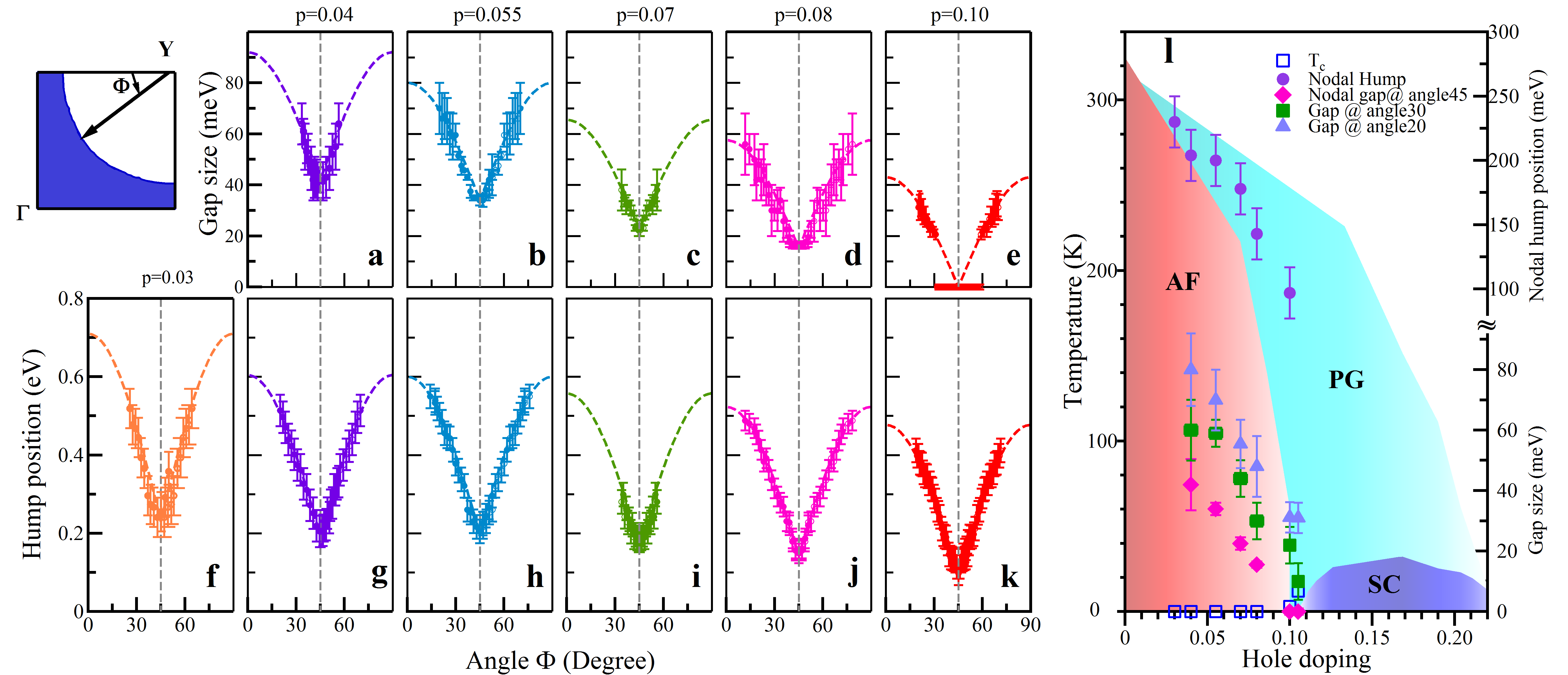}
\end{center}
\caption{{\bf Momentum dependence of the energy gap and its evolution with doping.} The data were taken at a temperature of 20K in lightly-doped La-Bi2201 samples.
({\bf a} to {\bf e}) show the momentum dependence of the energy gap with different doping levels {\it p} of 0.04 ({\bf a}), 0.055 ({\bf b}),
0.07 ({\bf c}), 0.08 ({\bf d}) and 0.10 ({\bf e}). The {\it $\Phi$} angle is defined in the top-left inset with $\Phi$=45 degrees corresponding
to the nodal direction. The gaps are symmetrized with respect to the nodal direction for better visualization (empty circles).  The dashed lines
depict the offset {\it d}-wave form $\Delta$=$\Delta_0$cos(2$\Phi$)+$\Delta_{\rm N}$ with the offset $\Delta_{\rm N}$ corresponding to the nodal
gap. Error bars in ({\bf a} to {\bf e}) reflect the maximum deviation of the fitting process. ({\bf f} to {\bf k}) Momentum evolution of the hump
for La-Bi2201 samples with different doping levels {\it p} of 0.03 ({\bf f}), 0.04 ({\bf g}), 0.055 ({\bf h}), 0.07 ({\bf i}), 0.08 ({\bf j})
and 0.10 ({\bf k}). The dashed lines represent a form of $P_{\rm h}$=$P_{\rm h0}$cos(2$\Phi$)+$P_{\rm hN}$ with $P_{\rm hN}$ representing the
hump position along the nodal direction. Error bars in ({\bf f} to {\bf k}) account for the uncertainty in the determination of the high
energy hump position. ({\bf l}) Phase diagram of La-Bi2201 system. It shows the three-dimensional antiferromagnetic region defined
by {\it T}$_{\rm N}$, superconducting region defined by {\it T}$_{\rm c}$ and the pseudogap region defined by {\it T}$^*$.  {\it T}$_{\rm N}$
and {\it T}$^*$ are reproduced from the previous NMR measurements\cite{NMRGQZheng}. {\it T}$_{\rm c}$ is determined from our magnetization and
resistivity measurements.  The doping dependence of the nodal gap size (pink diamond) and the nodal hump position (blue circles) is plotted.
The energy gap away from the nodal direction, with $\Phi$=30 degrees (green solid squares) and 20 degrees (blue solid triangles) is also plotted.
We note that at {\it p}$\sim$0.10, the nodal gap goes to zero, the three-dimensional antiferromagnetism vanishes ({\it T}$_{\rm N}$ goes to zero),
and superconductivity starts to emerge. The error bars for the gap and the hump are defined in the same way as in ({\bf a} to {\bf e}) and ({\bf f} to {\bf k}), respectively. }
\end{figure}


\textbf{References}}
\begin{thebibliography}{99}

\bibitem{PLee} Lee, P. A. {\it et al.} Doping a Mott insulator: Physics of high temperature superconductivity. {\it Rev. Mod. Phys.} {\bf 78}, 17-85 (2006).
\bibitem{CCTsuei} Tsuei, C. C. $\&$ Kirtley, J. R. Pairing symmetry in cuprate superconductors. {\it Rev. Mod. Phys.} {\bf 72}, 969-1016 (2000).
\bibitem{TimuskReview} Timusk, T. $\&$ Statt, B. The pseudogap in high-temperature superconductors: an experimental survey. {\it Rep. Prog. Phys.} {\bf 62}, 61-122 (1999).
\bibitem{IBozovic} Bollinger, A. T. {\it et al.} Superconductor-insulator transition in La$_{2-x}$Sr$_x$CuO$_4$ at the pair quantum resistance. {\it Nature} {\bf 472}, 458-460 (2011).
\bibitem{CCOCSTM} Kohsaka, Y. {\it et al.} Visualization of the emergence of the pseudogap state and the evolution to superconductivity in a lightly hole-doped Mott insulator. {\it Nat. Phys.} {\bf 8}, 534-538 (2012).
\bibitem{NMRGQZheng} Kawasaki, S. {\it et al.} Carrier-concentration dependence of the pseudogap ground state of superconducting Bi$_2$(Sr$_{2-x}$La$_x$)CuO$_{6+\delta}$ revealed by $^{65,66}$Cu-nuclear magnetic resonance in very high magnetic fields. {\it Phys. Rev. Lett.} {\bf 105}, 137002 (2010).
\bibitem{Fukuzumi49} Fukuzumi, Y. {\it et al.} Universal superconductor-insulator transition and {\it T}$_{\rm c}$ depression in Zn-substituted high-{\it T}$_{\rm c}$ cuprates in the underdoped regime. {\it Phys. Rev. Lett.} {\bf 76}, 684-687 (1996).
\bibitem{Mott50} Mott, N. F. $\&$ Davis, E. A. {\it Electron Process in Non-Crystalline Materials.} Clarendon, Oxford (1979).
\bibitem{Fiory51} Fiory, A. T. {\it et al.} Dimensionality of localization in nonsuperconducting Bi$_{2+x}$Sr$_{2-y}$CuO$_{6\pm\delta}$ crystals. {\it Phys. Rev. B} {\bf 41}, 2627(R) (1990).
\bibitem{UnderlyFS} Gros, C. {\it et al.} Determining the underlying Fermi surface of strongly correlated superconductors, {\it Proc. Natl Acad. Sci. USA} {\bf 103}, 14298-14301 (2006).
\bibitem{NormanEDC} Norman, M. R. {\it et al.} Phenomenology of the low-energy spectral function in high-{\it T}$_{\rm c}$ superconductors. {\it Phys. Rev. B} {\bf 57}, R11093-R11096 (1998).
\bibitem{NormanArc} Norman, M. R. {\it et al.} Destruction of the Fermi surface in underdoped high-{\it T}$_{\rm c}$ superconductors. {\it Nature} {\bf 392}, 157-160 (1998).
\bibitem{KanigelNodalLiquid} Kanigel, A. {\it et al.} Evolution of the pseudogap from Fermi arcs to the nodal liquid.  {\it Nat. Phys.} {\bf 2}, 447-451 (2006).
\bibitem{WSLee} Lee, W. S. {\it et al.} Abrupt onset of a second energy gap at the superconducting transition of underdoped Bi2212. {\it Nature} {\bf 450}, 81-84 (2007).
\bibitem{TYoshida} Yoshida, T. {\it et al.} Metallic behavior of lightly doped La$_{2-x}$Sr$_x$CuO$_4$ with a Fermi surface forming an arc. {\it Phys. Rev. Lett.} {\bf 91}, 027001 (2003).
\bibitem{KShenPRL} Shen, K. M. {\it et al.} Missing quasiparticles and the chemical potential puzzle in the doping evolution of the cuprate superconductors. {\it Phys. Rev. Lett.} {\bf 93}, 267002 (2004).
\bibitem{HashimotoPRB} Hashimoto, M. {\it et al.} Doping evolution of the electronic structure in the single-layer cuprate Bi$_2$(Sr$_{2-x}$La$_x$)CuO$_{6+\delta}$: Comparison with other single-layer cuprates. {\it Phys. Rev. B} {\bf 77}, 094516 (2008).
\bibitem{HDingPRB} Pan, Z. H. {\it et al.} Evolution of Fermi surface and normal-state gap in the chemically substituted cuprates Bi$_2$Sr$_{2-x}$Bi$_x$CuO$_{6+\delta}$. {\it Phys. Rev. B} {\bf 79}, 092507 (2009).
\bibitem{FRonning} Ronning, F. {\it et al.} Photoemission evidence for a remnant Fermi surface and a {\it d}.-wave-like dispersion in insulating Ca$_2$CuO$_2$Cl$_2$. {\it Science} {\bf 282}, 2067-2072 (1998).
\bibitem{Chatterjee} Chatterjee, U. {\it et al.} Observation of a {\it d}-wave nodal liquid in highly underdoped Bi$_2$Sr$_2$CaCu$_2$O$_{8+\delta}$. {\it Nat. Phys.} {\bf 6}, 99-103 (2010).
\bibitem{IVishik} Vishik, I. M. {\it et al.} Phase competition in trisected superconducting dome. {\it Proc. Natl Acad. Sci. USA} {\bf 6}, 18332-18337 (2012).
\bibitem{LSCONGap} Razzoli, E. {\it et al.} Evolution from a nodeless gap to {\it d}$_{x^2-y^2}$ form in underdoped La$_{2-x}$Sr$_x$CuO$_4$. {\it Phys. Rev. Lett.} {\bf 110}, 047004 (2013).
\bibitem{SLupi} Lupi, S. {\it et al.} Far-Infrared Absorption and the Metal-to-Insulator Transition in Hole-Doped Cuprates. {\it Phys. Rev. Lett.} {\bf 102}, 206409 (2009).
\bibitem{HTakagi} Takagi, H. {\it et al.} Systematic evolution of temperature-dependent resistivity in La$_{2-x}$Sr$_x$CuO$_4$. {\it Phys. Rev. Lett.} {\bf 69}, 2975-2978 (1992).
\bibitem{YAndo} Ando, Y. {\it et al.} Mobility of the Doped Holes and the  Antiferromagnetic Correlations in Underdoped High-{\it T}$_{\rm c}$ Cuprates. {\it Phys. Rev. Lett.} {\bf 87}, 017001 (2001).
\bibitem{KShenNGap} Shen, K. M. {\it et al.} Fully gapped single-particle  excitations in lightly doped cuprates. {\it Phys. Rev. B} {\bf 69}, 054503 (2004).
\bibitem{HBYang} Yang, H. B. {\it et al.} Reconstructed Fermi surface of underdoped Bi$_2$Sr$_2$CaCu$_2$O$_{8+\delta}$ cuprate superconductors, {\it Phys. Rev. Lett.} {\bf 107}, 047003 (2011).
\bibitem{XFSun} Sun, X. F. {\it et al.} Doping dependence of phonon and quasiparticle heat transport of pure and Dy-doped Bi$_2$Sr$_2$CaCu$_2$O$_{8+\delta}$ single crystals, {\it Phys. Rev. B} {\bf 77}, 094515 (2008).
\bibitem{HMatsui} Matsui, H. {\it et al.} Angle-resolved photoemission spectroscopy of the antiferromagnetic superconductor Nd$_{1.87}$Ce$_{0.13}$CuO$_4$: anisotropic spin-correlation gap, pseudogap, and the induced quasiparticle mass enhancement. {\it Phys. Rev. Lett.} {\bf 94}, 047005 (2005).
\bibitem{Atkinson} Atkinson, W. A. {\it et al.} Robustness of the nodal {\it d}-wave spectrum to strongly fluctuating competing order. {\it Phys. Rev. Lett.} {\bf 109}, 267004 (2012).
\bibitem{ESEffect} Efros, A. L. $\&$ Shklovskii, B. I. Coulomb gap and  low temperature conductivity of disordered systems. {\it J. Phys. C} {\bf 8}, L49-L51 (1975);
   Shklovskii, B. I. $\&$  Efros, A. L. Electronic Properties of Doped Semiconductors, Springer: New York (1984).
\bibitem{ZHPan} Pan, Z. H. {\it et al.} Evolution of Fermi surface and normal-state gap in the chemically substituted cuprates Bi$_2$Sr$_{2-x}$Bi$_x$CuO$_{6+\delta}$. {\it Phys. Rev. B} {\bf 79}, 092507 (2009).
\bibitem{XYZhang} Zhang, X. Y. {\it et al.} Mott transition in the {\it d=$\infty$} Hubbard model at zero temperature. {\it Phys. Rev. Lett.} {\bf 70}, 1666-1669 (1993).
\bibitem{DisorderNP} Bouadim, K. {\it et al.} Single- and two-particle energy gaps across the disorder-driven superconductor-insulator transition. {\it Nat. Phys.} {\bf 7}, 884-889 (2011).
\bibitem{LSCOstripe1} Wakimoto, S. {\it et al.} Observation of incommensurate magnetic correlations at the lower critical concentration
for superconductivity in La$_{2-x}$Sr$_x$CuO$_4$ ({\it x}=0.05) {\it Phys. Rev. B} {\bf61}, 3699-3706 (2000).
\bibitem{LSCOstripe2}Fujita, M. {\it et al.} Static magnetic correlations near the insulating-superconducting phase boundary in La$_{2-x}$Sr$_x$CuO$_4$. {\it Phys. Rev. B} {\bf65}, 064505 (2002).
\bibitem{Bi2201stripe} Enoki, M. {\it et al.} Spin-Stripe Density Varies Linearly With the Hole Content in Single-Layer Bi$_{2-x}$Sr$_{2+x}$CuO$_{6+y}$ Cuprate Superconductors. {\it Phys. Rev. Lett.} {\bf110}, 017004 (2013).

\bibitem{Valla2012} Valla, T. Angle-resolved photoemission from cuprates with static stripes. {\it Physica C} {\bf481}, 66-74 (2012).
\bibitem{Vallascience} Valla, T. {\it et al.} The ground state of the pseudogap in cuprate superconductors. {\it Science} {\bf 314}, 1914-1916 (2006).
\bibitem{PerfettiPolaron} Perfetti, L. {\it et al.} Mobile small polarons and the Peierls transition in the quasi-one-dimensional conductor K$_{0.3}$MoO$_3$. {\it Phys. Rev. B} {\bf 66}, 075107 (2002).
\bibitem{Mannella} Mannella, N. {\it et al.} Nodal quasiparticle in pseudogapped colossal magnetoresistive manganites. {\it Nature} {\bf 447}, 474-478 (2005).
\bibitem{ORosch} R\"{o}sch, O. {\it et al.} Polaronic behavior of undoped High-{\it T}$_{\rm c}$ cuprate superconductors from angle-resolved photoemission spectra. {\it Phys. Rev. Lett.} {\bf 95}, 227002 (2005).
\bibitem{XJZhouReview} Zhou, X. J. {\it et al.} in {\it Handbook of High-Temperature Superconductivity: Theory and Experiment}, Chapter 3, edited by J. R. Schrieffer and J. Brooks, (Springer, New York) (2007).
\bibitem{Sangiovanni} Sangiovanni, G. {\it et al.} Electron-phonon interaction and antiferromagnetic correlations. {\it Phys. Rev. Lett.} {\bf97}, 046404 (2006).
\bibitem{Mishchenko} Mishchenko, A. $\&$ Nagaosa, N. Electron-phonon coupling and a polaron in the t-J model: From the weak to the strong coupling regime. {\it Phys. Rev. Lett.} {\bf 93}, 036402 (2004).
%%\bibitem{XJZhouNature} Zhou, X. J. {\it et al.} Universal nodal Fermi velocity. {\it Nature} {\bf 423}, 398 (2003).
\bibitem{Bonca} Bonca, J. {\it et al.} Spectral properties of a hole coupled to optical phonons in the generalized t-J model. {\it Phys. Rev. B} {\bf 77}, 054519 (2008).
\bibitem{RVBTheory} Anderson, P. W. The resonating valence bond state in La$_2$CuO$_4$ and superconductivity. {\it Science} {\bf 235}, 1196-1198 (1987).
\bibitem{Lee} Lee, T. K., Ho, C. M. $\&$ Nagaosa, N. Theory for slightly doped antiferromagnetic Mott insulators. {\it Phys. Rev. Lett.} {\bf90}, 67001 (2003).
\bibitem{JQMengCrystal} Meng, J. Q. {\it et al.} Growth, characterization and physical properties of high-quality large single crystals of Bi$_2$(Sr$_{2-x}$La$_x$)CuO$_{6+\delta}$ high-temperature superconductors. {\it Supercond. Sci. Technol.} {\bf 22}, 045010 (2009).
\bibitem{OnoPRB} Ono, S. $\&$ Ando, Y. Evolution of the resistivity anisotropy in Bi$_2$Sr$_{2-x}$La$_x$CuO$_{6+\delta}$ single crystals for a wide range of hole doping. {\it Phys. Rev. B} {\bf 67}, 104512 (2003).
\bibitem{GDLiu} Liu, G. D. {\it et al.} Development of a vacuum ultraviolet laser-based angle-resolved photoemission system with a superhigh energy resolution better than 1 meV. {\it Rev. Sci. Instrum.} {\bf 79}, 023105 (2008).

\end{thebibliography}
\end{document}